\documentclass[runningheads]{llncs}
\usepackage{comment} 
\usepackage{graphicx} 
\usepackage{amsmath} 
\usepackage{amssymb} 
\usepackage{booktabs} 
\usepackage{cite} 
\usepackage{url}
\usepackage{subfigure}

\begin{document}

\title{A Transformer-based Network for Deformable Medical Image Registration}

\author{Yibo Wang \and Wen Qian \and Mengqi Li \and Xuming Zhang \thanks{Corresponding author: Xuming Zhang}} 

\authorrunning{Wang et al.}

\institute{Department of Biomedical Engineering,  College of Life Science and Technology, Huazhong University of Science and Technology, Wuhan, China \email{zxmboshi@hust.edu.cn} } 

\maketitle              % typeset the header of the contribution

\begin{abstract}
Deformable medical image registration plays an important role in clinical diagnosis and treatment. Recently, the deep learning (DL) based image registration methods have been widely investigated  and showed excellent performance in computational speed. However, these methods cannot provide enough registration accuracy because of insufficient ability in representing both the global and local features of the moving and fixed images. To address this issue, this paper has proposed the transformer based  image registration method. This method uses the distinctive transformer to extract the global and local image features for generating the deformation fields, based on which the registered image is produced in an unsupervised way. Our method can improve the registration accuracy effectively by means of self-attention mechanism and bi-level information flow. Experimental results on such brain MR image datasets as LPBA40 and OASIS-1 demonstrate that compared with several traditional and DL based registration methods, our method provides higher registration accuracy in terms of dice values.


\end{abstract}

\section{Introduction}
\label{sec:introduction}

Image registration is one of the fundamental and challenging tasks in medical image processing and analysis. Its goal is to find the correspondence between the moving and fixed images to facilitate such tasks as disease diagnosis and surgical navigation.  Up to now, the various image registration methods have been proposed. For the traditional registration methods \cite{2007A,2008Symmetric,2010elastix}, the similarity metric is firstly constructed between the fixed and  moving images. Then,  the objective function based on the constructed metric is optimized to produce the registered image. These methods are time-consuming because of the complicated iterative optimization. 

To improve image registration efficiency, the deep learning (DL) based registration methods have been presented. Given numerous moving and fixed images, the deep neural networks can be trained to generate the registered image efficiently. Depending on how the networks are trained, these methods can be categorized into the supervised learning and unsupervised learning ones. In the supervised approaches, the ground-truth deformation fields or anatomical landmarks are needed \cite{miao2016real, 2017Nonrigid, 2017Quicksilver, Marc2017SVF, 2018Deep, salehi2018real}. Sokooti et al. \cite{2017Nonrigid} have proposed a convolution neuron network (CNN) to directly estimate the displacement vector field (DVF) using the artificially generated DVFs. Cao et al. \cite{2018Deep} have developed a deformable inter-modality image registration method which estimates the deformation fields using the deep neural network supervised by intra-modality similarity. The registration performance of these methods greatly depends on the ground-truths which are generally difficult to acquire in clinical scenarios. As for the unsupervised learning based methods \cite{vos2017end, yoo2017ssemnet, 2018An, 2020Recursive, 20204D, 2021CycleMorph}, they need no the ground truth of the deformation field. Balakrishnan et al. \cite{2018An} have proposed a 3D medical image registration model, voxelmorph, which reconstructs the registered result using a CNN with a spatial transform layer. Zhao et al. \cite{2020Recursive} have designed a volume tweening network (VTN) including the cascaded subnetworks  to improve the registration performance recursively. Kim et al. \cite{2021CycleMorph} have presented a cycle-consistent deformable image registration method called cyclemorph, which can  enhance the registration performance by introducing the cycle consistency into the network loss.

Although the existing DL based registration approaches can provide higher computational efficiency than the traditional ones, they cannot capture the long-range dependence in the moving and the fixed image effectively because of the adoption of such networks as the CNN which has the limited ability of extracting the global image features. Therefore, the registration accuracy of these DL based methods is influenced disadvantageously especially when the large deformation is involved between the fixed and moving images.  Recently, the transformer has become an important network in the fields of natural language processing and computer vision because it can explore the long-range dependence based on the self-attention mechanism. Distinctively, the transformer can extract the global image features effectively, thus it has been applied to such tasks  as image classification \cite{2020An, he2022fully}, image denoising \cite{2021Eformer, wang2021global} and image segmentation \cite{chen2021transunet, cao2021swin, ma2022ht}.  To overcome the disadvantages of existing CNN based registration methods,  we have presented a novel deformable medical image registration network called Transformer-UNet (TUNet). This network introduces the vision transformer (VIT) \cite{2020An} into the framework of UNet \cite{2015U} to extract the global and local features from the moving and fixed images, thereby generating the deformation field effectively.  Besides,  the skip connections are established in the bi-level layers to guarantee the correct information flow between the rough features and the fine features. 

Experiments have been done on LPBA40 and OASIS-1 to test the performance of our method. The qualitative and quantitative evaluations demonstrate that  the proposed method is provided with higher registration accuracy than the compared traditional and DL based registration methods. 

The paper is organized as follows. Section \ref{sec:method} describes our method. Section \ref{sec:experiment} presents the experimental results of our method on two datasets. Conclusion is given in Section \ref{sec:conclusion}.

\section{Method} 
\label{sec:method}

\begin{figure*}[hbp]
	\centering
	\includegraphics[width=1\textwidth]{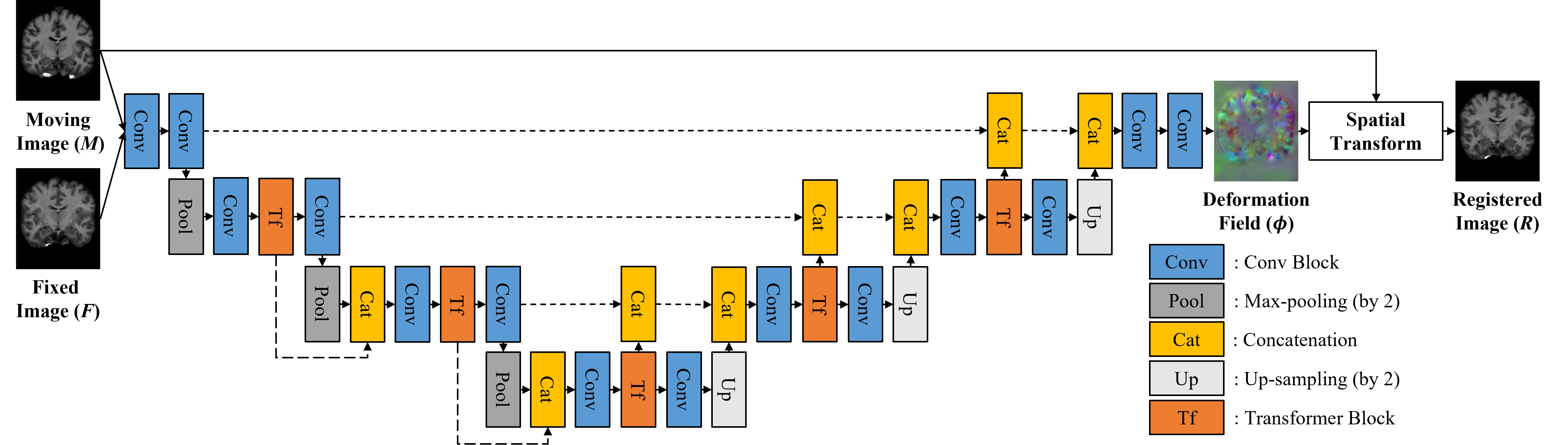}
	% \vspace{-1.8em}
	\caption{The overall framework of the proposed method, Transformer-UNet, for deformable medical image registration. Here, the short and long dashed lines denote the skip connection and the bi-level connection, respectively. }
	\label{fig:workFlow}
\end{figure*}

The framework of the TUNet is shown in Fig.  \ref{fig:workFlow}. Here,  a moving image $M$ and a fixed image $F$ are input into the TUNet. The deformation field $\phi$ is computed based on the parameters learned in the different network layers. By means of the spatial transform layer,   $M$ is deformed to produce the registered image $R$.  The TUNet is trained using the loss defined by the dissimilarity between $F$  and $M$ and the smoothness constraint of $\phi$ to produce the registered result in an unsupervised way.

\subsection{Architecture of the Transformer-UNet}

Our Transformer-UNet is built on the encoder-decoder architecture of the UNet \cite{2015U}, but improves the latter by introducing the bi-level connection and an unique Transformer block. As shown in Fig. \ref{fig:workFlow}, the proposed Transformer-UNet uses a single input formed by concatenating $M$ and $F$ in the dimension channel. In the encoder, the two Conv layers are used  to extract image features, where each block is composed of a convolutional module followed by the Rectified Linear Unit (ReLU). The kernel size and stride in the convolutional module will be set to 3$\times$3$\times$3 and 1, respectively. The Max-pooling layer, Conv layers and Transformer blocks are combined to produce image features at different levels. In some layers in the encoder, the concatenation layer is additionally introduced to concatenate the features produced by the Transformer block at the previous layer and those resulting from the pooling layer. In the decoder,  the Conv layers, Transformer blocks and Up-sampling layer are combined to store the spatial resolution of the feature maps at different levels.  The features at the same levels produced by the encoder and decoder will be concatenated by the concatenation layers. At the end of decoder, the concatenated features will be processed by the two Conv layers to output the final features. Note that the stride of Max-pooling layer and Up-sampling layer is set to 2, and thus the encoder reduces the spatial resolution of input volumes by a factor of 8 in total and the decoder restores the features to the original size. As the key component of our method, the Transformer block is distributed at different layers in the encoder and decoder. It receives the convolutional features and outputs two different feature maps. 

\subsection{Transformer Block}

\begin{figure}
	\centering
	\includegraphics[width=0.8\textwidth]{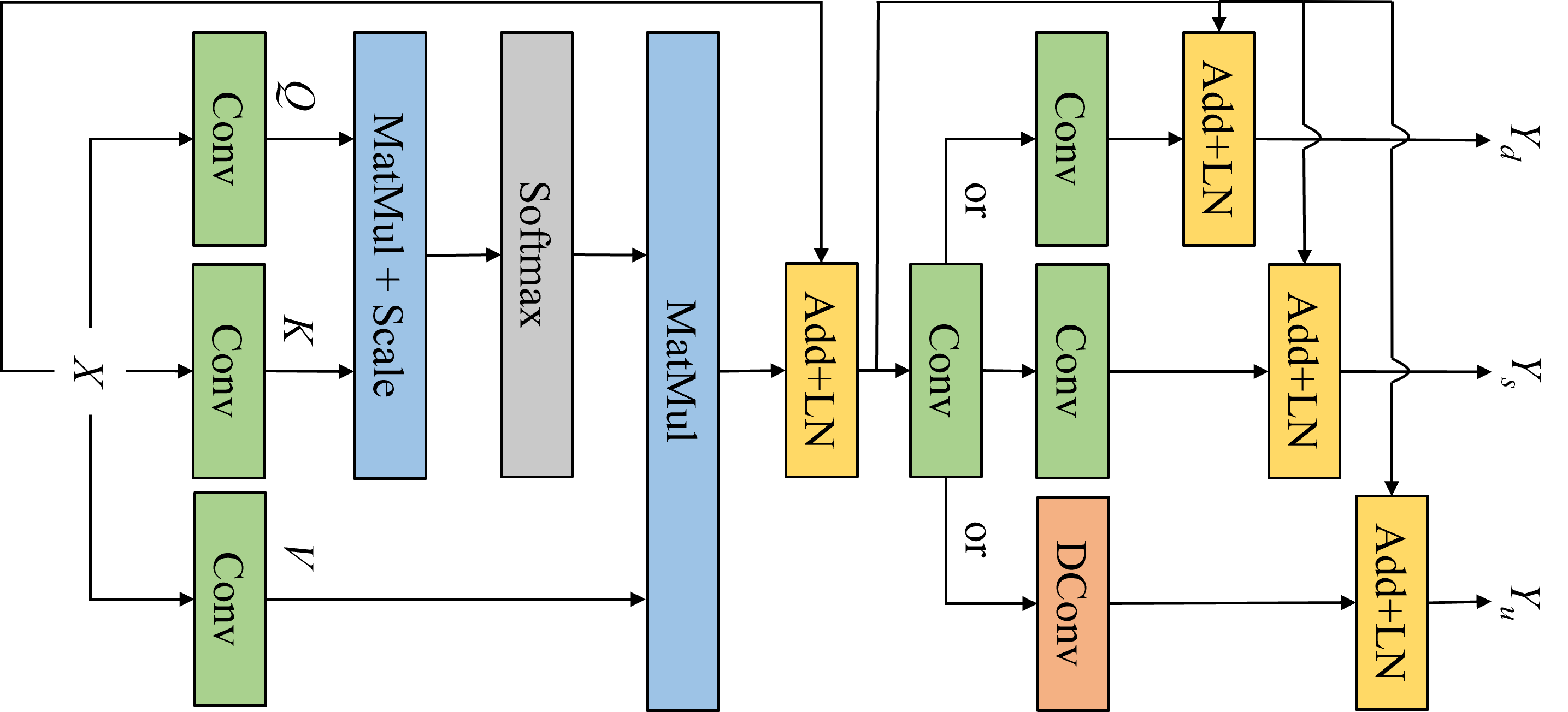}
	% \vspace{-0.5em}
	\caption{The structure of the proposed Transformer block.}
	\label{fig:transBlock}
	% \vspace{-1.2em}
\end{figure}

Inspired by the VIT \cite{2020An}, we will build the Transfomer block shown in Fig. \ref{fig:transBlock}. Compared with the VIT, this block retains the multi-head self-attention mechanism, which is necessary for improving network's awareness for the global information. Meanwhile, we have made some modifications on the VIT to produce the distinctive Transfomer block. Firstly,  the redundant position embedding is removed after the patch embedding. Secondly, the convolution module is used to directly compute the weight matrix instead of the original patch embedding and the linear mapping, which will reduce the computational complexity and  meet the need of 3D image registration better. Finally,  an additional output path has been designed. By using the convolution module with a stride of 2 or the deconvolution module, we have established a bridge for information flow in the bi-level features. 

As shown in Fig. \ref{fig:transBlock}, the input feature $X$ is processed by the three different convolution modules to generate the query matrix $Q$, the key matrix $K$ and the value matrix $V$  as: 
% \vspace{-0.3em}
\begin{equation}
Q = W_Q \cdot X; K = W_K \cdot X; V = W_V \cdot X,
\end{equation}

The matrices $Q$, $K$ and  $V$ are reshaped into a sequence of flattened 3D patches: $Q$, $K$, and $V$ $\in$ $\mathbb{R}^{B \times N \times (P^3 \cdot C)}$, where $B$ is the mini-batch, $C$ is the number of channels, $(P, P, P)$ is the resolution of each volume patch, and $N=HWD/P^3$ is the resultant number of patches with $(H, W, D)$ denoting the resolution of the input feature. By splitting the heads from the embed channels and swapping the order of axis, we will change them into a 4D vector: $Q$, $V$ $\in$ $\mathbb{R}^{B \times k \times N \times d_k}$, $K$ $\in$ $\mathbb{R}^{B \times k \times d_k \times N}$, where $k$ is the number of heads and $d_k=(P^3 \cdot C) / k$ is the number of the embedding channels per head. The matrix multiplication, scaling and softmax operations will be implemented for $Q$, $K$ and $V$ to produce the output $Y$. 
% \vspace{-0.5em}
\begin{equation}
Y = softmax(\frac{QK^T}{\sqrt{d_k}})V,
\end{equation}

The output $Y$ will be added to the input $X$ and then is processed by the layer-norm (LN) operation. To promote the information interaction between two different levels of features, we will add the convolution module with a stride of 2 or the deconvolution (DConv) module at the end of block. Similarly, the LN will be applied again. In this way, our Transformer block will generate two output feature maps. One called $Y_s$ has the same size as the input feature map while another $Y_d$ or $Y_u$ has half or twice the size of the input feature map. 

\subsection{Spatial Transform}
For the spatial transform, we will choose the 3D transformation function with the bilinear interpolation defined as: 
% \vspace{-0.6em}
\begin{equation}
M \circ \phi = \sum\limits_{q \in G(\phi(p))} M(q) \prod\limits_{d \in \left\{ x, y, z \right\} } (1 - \left| \phi_d(p) - q_d \right|),
\end{equation}
where $p$ is a voxel,  $G(\phi(p))$ means the 8-neighbors of $\phi(p)$ and $\circ$ is the spatial transform function. 

\subsection{Loss Function}
The loss function of the TUNet includes a dissimilarity term related to the local cross correlation ($CC$) and a smoothness regularization term of $\phi$, and it is defined as: 
% \vspace{-0.5em}
\begin{equation}
\mathcal{L}(M, F, \phi) = -CC(M \circ \phi, F) + \lambda \sum\limits_{p \in \Omega} \left \| \nabla \phi(p) \right \|,
\end{equation}
where $\lambda$ is a hyper-parameter and $\Omega$ denotes the 3D volume and $CC(A, B)$ is computed as:
% \vspace{-0.5em}
\begin{equation}
CC(A, B)\! =\!\sum\limits_{v \in \Omega} \frac{(\sum_{v_i}(A(v_i) - \bar{A}(v)) (B(v_i) - \bar{B}(v)))^2}{\sum_{v_i}(A(v_i) - \bar{A}(v))^2\sum_{v_i}(B(v_i) - \bar{B}(v))^2},
\end{equation}
where $v_i$ is chosen as the $9\times9\times9$ patch, $\bar{A}(v)$ and $\bar{B}(v)$ mean the local mean of  $A(v)$ and $B(v)$, respectively.  

\section{Experimental Results and Discussion}
\label{sec:experiment}

% 实验设置
\subsection{Experimental Settings}

% 数据集 
\subsubsection{Datasets} We have chosen LPBA40 \cite{2008Construction} and OASIS-1 \cite{Daniel2007Open} for experiments. The LPBA40 contains 40 T1-weighted brain MR images, where  56 anatomical areas are segmented from each image. The OASIS-1 contains 414 T1-weighted brain MR images, where each image includes 35 segmented cortical regions. Here, all scans are sampled to a 256$\times$256$\times$256 grid with 1mm isotropic voxel. The affine spatial normalization and brain extraction are carried out using FreeSurfer \cite{2012FreeSurfer}. The images are further cropped into 192$\times$160$\times$192. For the LPBA40, we have trained our model on 25 subjects, validated it on 5 subjects and tested it on 5 subjects. For the OASIS-1, we have used 324 subjects for model training, 42 subjects for validation and 40 subjects for testing. What's more, we use random rotation and flipping to form 12 $\times$ data augmentation. 

% 实施细节 
\subsubsection{Implementation Details} We will focus on atlas-based registration, in which  a fixed volume is chosen as atlas and  each volume in the dataset is registered to it. Here, because of high memory cost in the training stage, we will extract patches of size 128$\times$128$\times$64 from a whole volume, and set the corresponding batch size according to the GPU memory usage. To avoid over-fitting, the random rotation is implemented on each training volume pair to realize data augmentation. We set hyper-parameter $\lambda$ to 0.1 and adopt Adam optimization with a learning rate of 1e-4. Our model is realized using the MindSpore Lite tool \cite{MindSpore} and it trained for 30 epochs on a single NVIDIA RTX 2080Ti GPU.

% 对比算法
\subsubsection{Compared Methods} In order to verify the superiority of the TUNet, we will compare it with several popular  image registration methods including  SyN \cite{2008Symmetric} from Advanced Normalization Tools, VoxelMorph (VM) \cite{2018An}, VTN \cite{2020Recursive}, CycleMorph (CM) \cite{2021CycleMorph}, TransUnet (TF1) \cite{chen2021transunet} and SwinUnet (TF2) \cite{cao2021swin}. As regards the VoxelMorph,  we will choose VoxelMorph-1 \cite{2018An} as the baseline network with the same parameters to our method for the fair comparison. Since TransUnet \cite{chen2021transunet} and SwinUnet \cite{cao2021swin} are image segmentation methods using transformer thinking, we rewrite them as registration method while retaining their structure. 

% 衡量指标
\subsubsection{Evaluation Metrics} The registration performance is evaluated by Dice  \cite{dice1945measures}, which is defined as the overlap rate between the segmented results of registered and fixed images. 

\begin{equation}
Dice(R, F) = 2 \cdot \frac{\left| R \cap F \right|}{\left| R \right| + \left| F \right|}
\end{equation}

Perfectly overlapped regions come with a Dice value of 1. The Dice value explicitly measures the coincidence between two regions and thereby reflects the quality of registration. Considering multiple anatomical structures annotated, we compute the Dice score with respect to each and take an average.  

\subsection{Ablation Experiment} 
To investigate different components' individual contribution towards model's overall performance, we progressively integrate our main contributions (Transformer Block and Bi-level Connection) into the model. Corresponding results in Table \ref{tab:ablation} reveals the importance of introducing Transformer Block together with Bi-level Connection along with the basic UNet architecture. Data in Table \ref{tab:ablation} are composed of 56 regional Dice averages from all test samples on LPBA40. We can notice that all of these components contribute towards model's performance. We also observe intriguing accuracy improvement created by Transformer Block (2.5 \%). Compared with Transformer Block, Bi-level Connection's effect is less satisfactory in that it only brings a slight increase in Dice value (0.6 \%). 

\begin{table}
	\caption{Ablation experiment on LPBA40.}
	\begin{center}
		\begin{tabular}{lc}
			\toprule
			Methods & Dice \\
			\midrule
			UNet & 0.5747 \\ 
			UNet + TransBlock & 0.5895 \\
			UNet + TransBlock + BiLevelConnection & 0.5932  \\
			\bottomrule
		\end{tabular}
		\label{tab:ablation}
	\end{center}
\end{table}

\subsection{Computational Efficiency} 
Considering actual implementation of the model, we compares the computation efficiency of different methods according to Giga Floating-point Operations Per Second (GFLOPs) and inference time on CPU (Intel(R) Core(TM) i7-6950X 3.00GHz) and GPU (NVIDIA RTX 2080Ti). Our results in Table \ref{tab:computaion} suggest that the inference time of the SyN is much longer than that of the DL-based methods. On CPU, all DL-based methods can realize image registration in 30 seconds while the SyN requires more than 1800 seconds. If GPU is used, the implementation time of DL-based methods can be shortened to less than 1 second. Apart from running time, GFLOPs can also reflect the superiority of our method. Although the TUNet is more computationally expensive than the fully convolutional networks \cite{2018An, 2020Recursive}, it still outperforms other transformer-based methods \cite{chen2021transunet, cao2021swin}. 

\begin{table}
	\caption{Comparison of running time and FLOPs}
	\begin{center}
		\begin{tabular}{cccccccc}
			\toprule
			Metric & SyN & VM & VTN & CM & TF1 & TF2 & TUNet \\
			\midrule
			Time(CPU) & 1853.16 & 7.76 & 23.28 & 31.04 & 23.58 & 24.32 & 29.24 \\ 
			Time(GPU) & None & 0.15 & 0.44 & 0.61 & 0.75 & 0.86 & 0.57 \\
			GFLOPs & None & 97.44 & 292.32 & 389.76 & 418.77 & 472.31 & 296.07 \\
			\bottomrule
		\end{tabular}
		\label{tab:computaion}
	\end{center}
\end{table} 

\subsection{Registration Results} 

\subsubsection{Quantitative Evaluation}

\begin{figure}[htbp]
\centering
\subfigure[Boxplots of Dice values on the first 8 anatomical structures]{
	\includegraphics[width=1\textwidth, height=0.25\textheight]{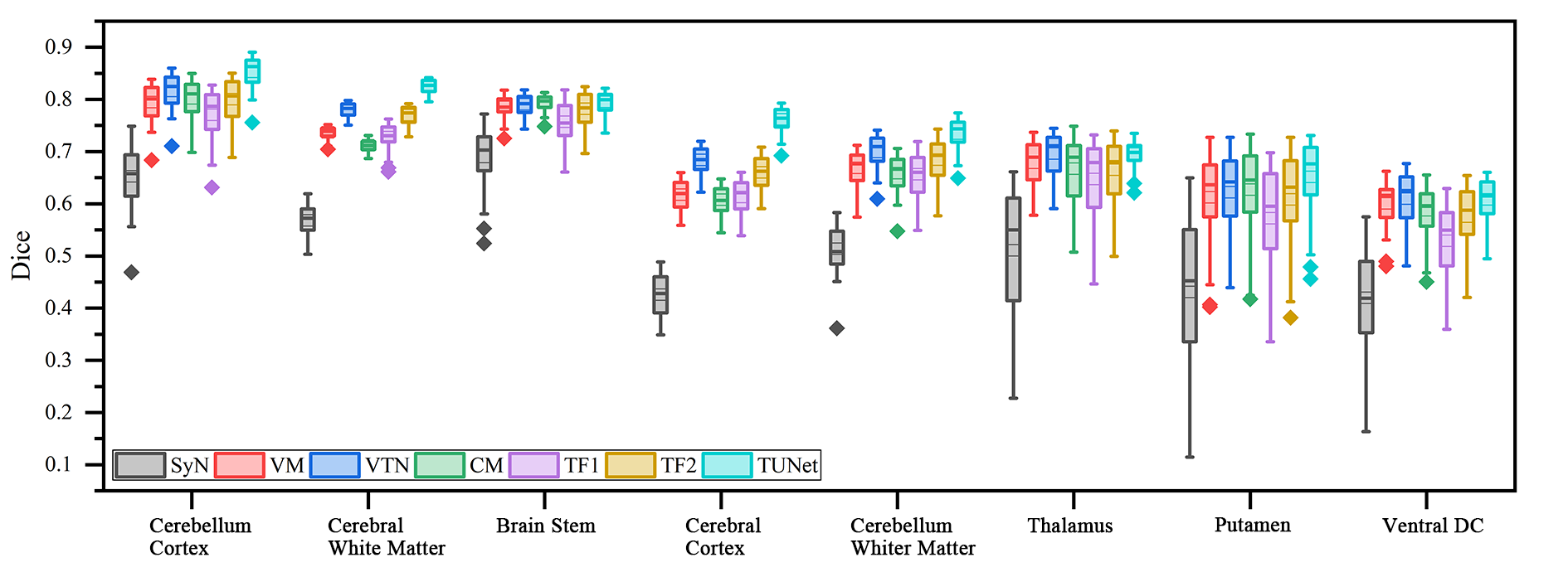}}
\subfigure[Boxplots of Dice values on the last 8 anatomical structures]{
	\includegraphics[width=1\textwidth, height=0.25\textheight]{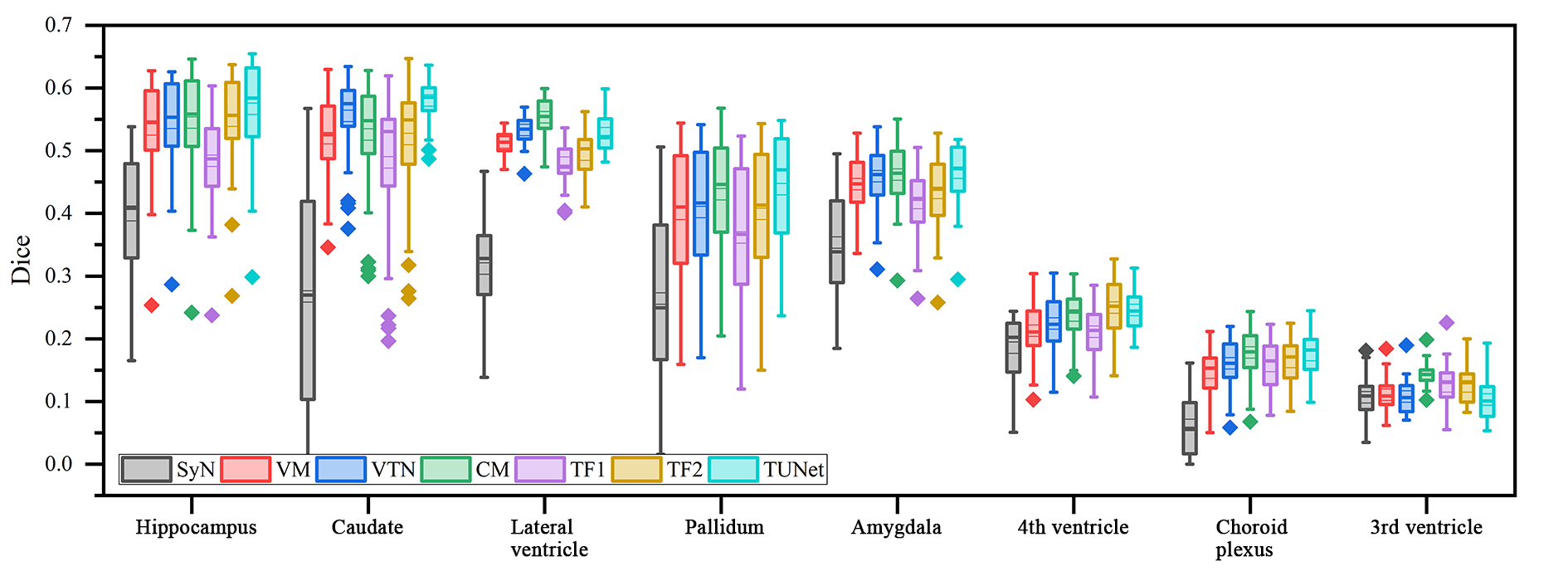}}
\caption{Boxplots of Dice values for the SyN, VTN, VM, CM, TF1, TF2 and TUNet performed on the anatomical structures. }
\label{fig:diceBoxplot}
\end{figure}

Fig. \ref{fig:diceBoxplot} visualizes the Dice values for 16 evaluated anatomical structures across all test samples on OASIS-1. For better visualization purpose, we combine the same structures from the left and right hemisphere together. For most of structures, our TUNet model achieves higher scores than VoxelMorph \cite{2018An} in short-range registration and TransUnet \cite{chen2021transunet} in long-range registration. In particular, on some structures such as Cerebellum cortex, Cerebral white matter, Cerebral cortex, Putamen and Hippocampus, our TUNet performs much better than the compared methods. As can be seen from Fig. \ref{fig:diceBoxplot}, the Dice values of our method exceed those of the fully convolutional models, which indicates the proposed Transformer Block's unique contribution. Compared with other methods with transformer, the proposed method has better registration performance, which proves that our designed bi-level skip connection is effective and irreplaceable. 

\vspace{-0.5cm}

\begin{figure}[htbp] 
\centering
\subfigure[The registered results of various methods (slice = 80)]{
\includegraphics[scale=0.65]{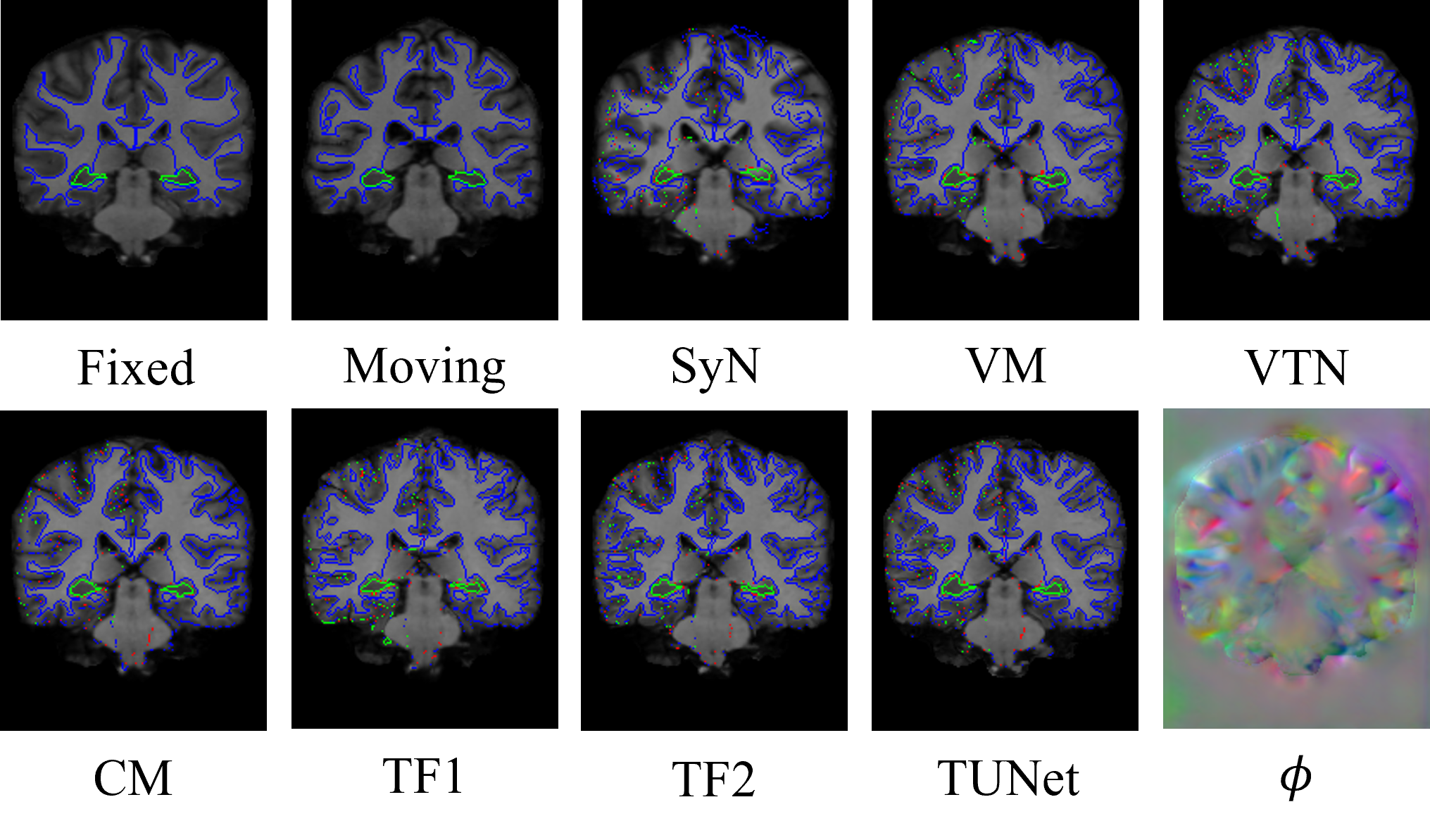}}

\subfigure[The registered results of various methods (slice = 100)]{
\includegraphics[scale=0.65]{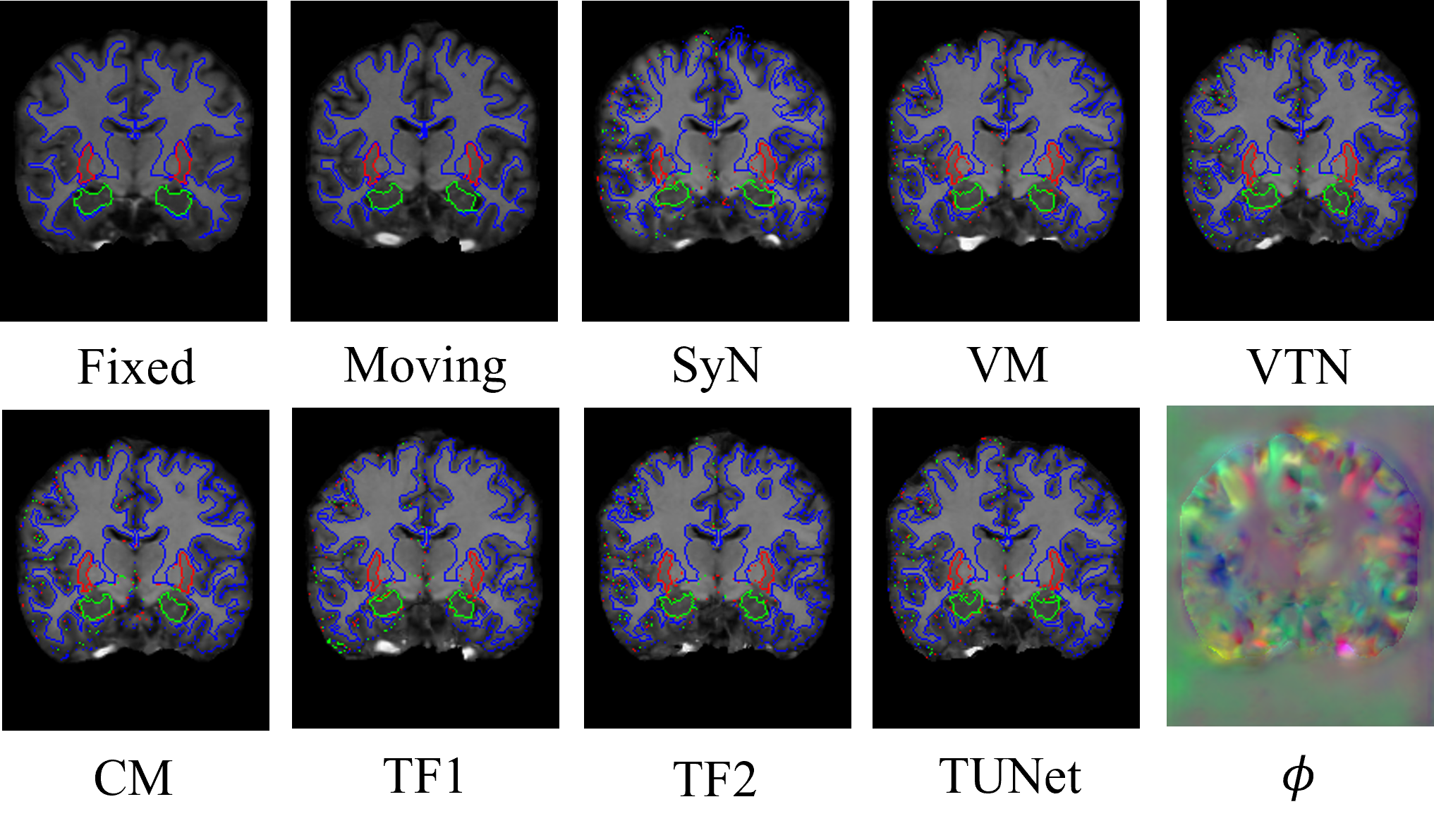}}

\subfigure[The registered results of various methods (slice = 120)]{
\includegraphics[scale=0.65]{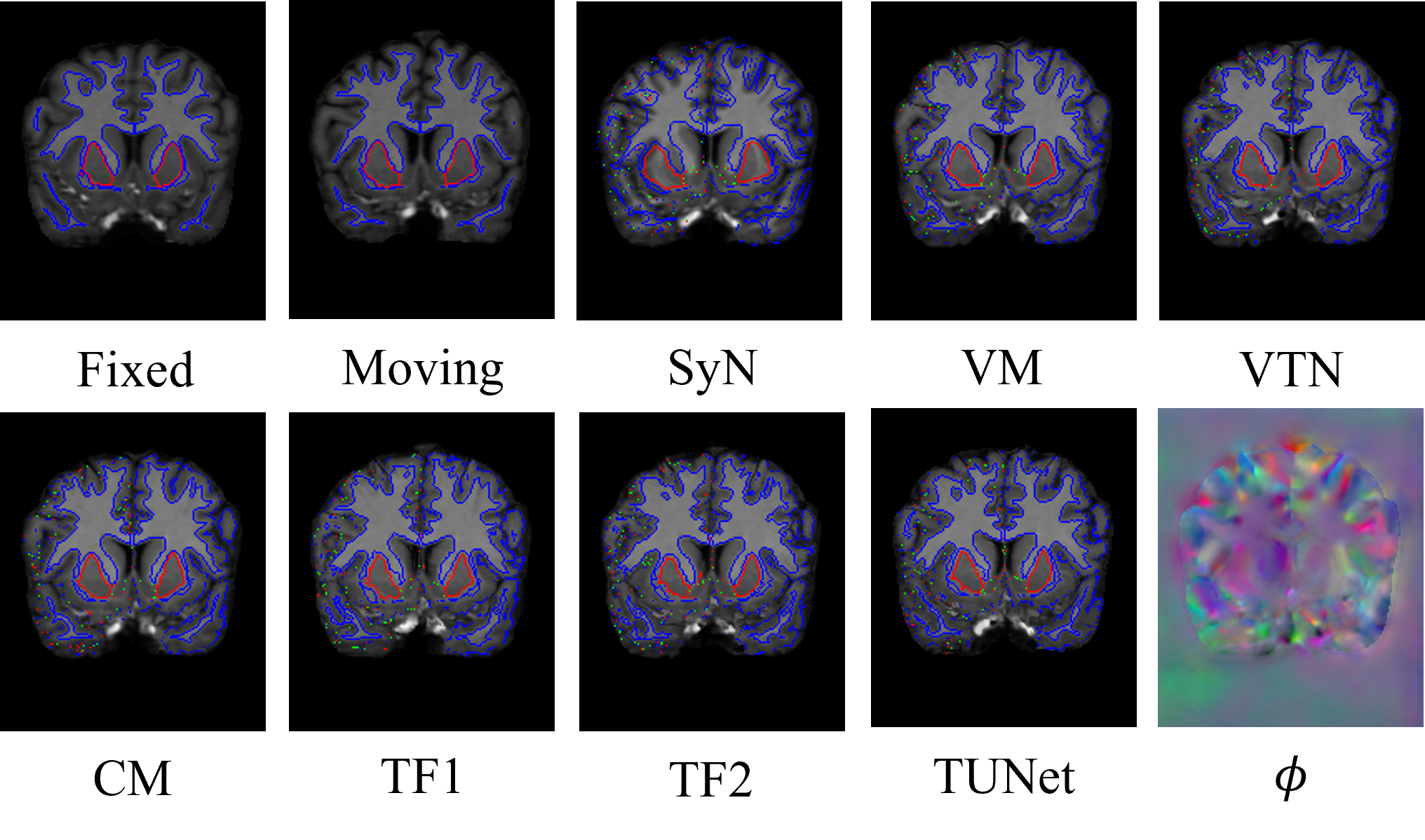}}

\caption{The registered results of the various methods on OASIS-1.}
\label{fig:sliceGroup}
\end{figure}

\subsubsection{Qualitative Evaluation}

The results of atlas-based OASIS-1 MR image registration are shown in Fig. \ref{fig:sliceGroup}. The proposed TUNet method can accurately register each pair of moving and fixed images, which can be specifically verified with the segmentation boundaries of several brain structures (blue outlines represent Cerebral White Matter, green outlines represent Hippocampus and red outlines represent Putamen respectively). Besides, the observation from Fig. \ref{fig:sliceGroup} shows that the other compared methods cannot restore the deformation of some internal brain structures and brain contour effectively. By comparison, the proposed TUNet can provide the most similar registered results to the fixed image in terms of the different brain tissues among all evaluated methods. Especially, the proposed method can preserve some fine image details better than the other methods, which can be verified with the red and green outlines. Indeed, the superiority of the TUNet to other methods lies in its outperforming ability to extract both the global and local features, which can be seen from the blue outlines and the  deformation field. 

\section{Conclusion} 
\label{sec:conclusion} 
In this paper, we have presented a Transformer-UNet based unsupervised deformable medical image registration method. The framework is built on the Transformer model which is introduced into the UNet. By means of  the distinctive Transformer-UNet,  the global and local features can be extracted from the moving and fixed images effectively, thereby ensuring good registration performance. The experimental results show that our method outperforms several traditional and deep learning based registration methods in terms of visual evaluations and such quantitative metrics as Dice. Future research will be focused on the extension of our method to multi-modal medical image registration.

\subsubsection{Acknowledgment}  
This work was sponsored by the National Natural Science Foundation of China (Grant No. 61871440) and CAAI-Huawei MindSpore Open Fund. We gratefully acknowledge the support of MindSpore.

\bibliographystyle{splncs04}
\bibliography{mybib}

\begin{thebibliography}{10}
\providecommand{\url}[1]{\texttt{#1}}
\providecommand{\urlprefix}{URL }
\providecommand{\doi}[1]{https://doi.org/#1}

\bibitem{2007A}
Ashburner, J.: A fast diffeomorphic image registration algorithm. Neuroimage
  \textbf{38}(1),  95--113 (2007)

\bibitem{2008Symmetric}
Avants, B.B., Epstein, C.L., Grossman, M., Gee, J.C.: Symmetric diffeomorphic
  image registration with cross-correlation: evaluating automated labeling of
  elderly and neurodegenerative brain. Medical Image Analysis  \textbf{12}(1),
  26--41 (2008)

\bibitem{2018An}
Balakrishnan, G., Zhao, A., Sabuncu, M.R., Dalca, A.V., Guttag, J.: An
  unsupervised learning model for deformable medical image registration. In:
  Proceedings of the IEEE/CVF Conference on Computer Vision and Pattern
  Recognition. pp. 9252--9260. IEEE (2018)

\bibitem{cao2021swin}
Cao, H., Wang, Y., Chen, J., Jiang, D., Zhang, X., Tian, Q., Wang, M.:
  Swin-unet: Unet-like pure transformer for medical image segmentation. arXiv
  preprint arXiv:2105.05537  (2021)

\bibitem{2018Deep}
Cao, X., Yang, J., Wang, L., Xue, Z., Wang, Q., Shen, D.: Deep learning based
  inter-modality image registration supervised by intra-modality similarity.
  In: International Workshop on Machine Learning in Medical Imaging. pp.
  55--63. Springer (2018)

\bibitem{chen2021transunet}
Chen, J., Lu, Y., Yu, Q., Luo, X., Adeli, E., Wang, Y., Lu, L., Yuille, A.L.,
  Zhou, Y.: Transunet: Transformers make strong encoders for medical image
  segmentation. arXiv preprint arXiv:2102.04306  (2021)

\bibitem{dice1945measures}
Dice, L.R.: Measures of the amount of ecologic association between species.
  Ecology  \textbf{26}(3),  297--302 (1945)

\bibitem{2020An}
Dosovitskiy, A., Beyer, L., Kolesnikov, A., Weissenborn, D., Zhai, X.,
  Unterthiner, T., Dehghani, M., Minderer, M., Heigold, G., Gelly, S., et~al.:
  An image is worth 16x16 words: Transformers for image recognition at scale.
  arXiv preprint arXiv:2010.11929  (2020)

\bibitem{2012FreeSurfer}
Fischl, B.: Freesurfer. Neuroimage  \textbf{62}(2),  774--781 (2012)

\bibitem{he2022fully}
He, X., Tan, E.L., Bi, H., Zhang, X., Zhao, S., Lei, B.: Fully transformer
  network for skin lesion analysis. Medical Image Analysis  \textbf{77},
  102357 (2022)

\bibitem{2021CycleMorph}
Kim, B., Kim, D.H., Park, S.H., Kim, J., Lee, J.G., Ye, J.C.: Cyclemorph: Cycle
  consistent unsupervised deformable image registration. Medical Image Analysis
   \textbf{71},  102036 (2021)

\bibitem{2010elastix}
Klein, S., Staring, M., Murphy, K., Viergever, M.A., Pluim, J.P.: Elastix: a
  toolbox for intensity-based medical image registration. IEEE Transactions on
  Medical Imaging  \textbf{29}(1),  196--205 (2009)

\bibitem{20204D}
Lei, Y., Fu, Y., Wang, T., Liu, Y., Patel, P., Curran, W.J., Liu, T., Yang, X.:
  4d-ct deformable image registration using multiscale unsupervised deep
  learning. Physics in Medicine \& Biology  \textbf{65}(8),  085003 (2020)

\bibitem{2021Eformer}
Luthra, A., Sulakhe, H., Mittal, T., Iyer, A., Yadav, S.: Eformer: Edge
  enhancement based transformer for medical image denoising. arXiv preprint
  arXiv:2109.08044  (2021)

\bibitem{ma2022ht}
Ma, M., Xia, H., Tan, Y., Li, H., Song, S.: Ht-net: hierarchical
  context-attention transformer network for medical ct image segmentation.
  Applied Intelligence pp. 1--14 (2022)

\bibitem{Daniel2007Open}
Marcus, D.S., Wang, T.H., Parker, J., Csernansky, J.G., Morris, J.C., Buckner,
  R.L.: Open access series of imaging studies (oasis): cross-sectional mri data
  in young, middle aged, nondemented, and demented older adults. Journal of
  Cognitive Neuroscience  \textbf{19}(9),  1498--1507 (2007)

\bibitem{miao2016real}
Miao, S., Wang, Z.J., Zheng, Y., Liao, R.: Real-time 2d/3d registration via cnn
  regression. In: 2016 IEEE 13th International Symposium on Biomedical Imaging
  (ISBI). pp. 1430--1434. IEEE (2016)

\bibitem{MindSpore}
MindSpore: https://www.mindspore.cn/

\bibitem{Marc2017SVF}
Roh{\'e}, M.M., Datar, M., Heimann, T., Sermesant, M., Pennec, X.: Svf-net:
  Learning deformable image registration using shape matching. In:
  International Conference on Medical Image Computing and Computer-assisted
  Intervention. pp. 266--274. Springer (2017)

\bibitem{2015U}
Ronneberger, O., Fischer, P., Brox, T.: U-net: Convolutional networks for
  biomedical image segmentation. In: International Conference on Medical Image
  Computing and Computer-assisted Intervention. pp. 234--241. Springer (2015)

\bibitem{salehi2018real}
Salehi, S.S.M., Khan, S., Erdogmus, D., Gholipour, A.: Real-time deep pose
  estimation with geodesic loss for image-to-template rigid registration. IEEE
  Transactions on Medical Imaging  \textbf{38}(2),  470--481 (2018)

\bibitem{2008Construction}
Shattuck, D.W., Mirza, M., Adisetiyo, V., Hojatkashani, C., Salamon, G., Narr,
  K.L., Poldrack, R.A., Bilder, R.M., Toga, A.W.: Construction of a 3d
  probabilistic atlas of human cortical structures. Neuroimage  \textbf{39}(3),
   1064--1080 (2008)

\bibitem{2017Nonrigid}
Sokooti, H., De~Vos, B., Berendsen, F., Lelieveldt, B.P., I{\v{s}}gum, I.,
  Staring, M.: Nonrigid image registration using multi-scale 3d convolutional
  neural networks. In: International Conference on Medical Image Computing and
  Computer-assisted Intervention. pp. 232--239. Springer (2017)

\bibitem{vos2017end}
Vos, B.D.d., Berendsen, F.F., Viergever, M.A., Staring, M., I{\v{s}}gum, I.:
  End-to-end unsupervised deformable image registration with a convolutional
  neural network. In: Deep learning in medical image analysis and multimodal
  learning for clinical decision support, pp. 204--212. Springer (2017)

\bibitem{wang2021global}
Wang, Z., Xie, Y., Ji, S.: Global voxel transformer networks for augmented
  microscopy. Nature Machine Intelligence  \textbf{3}(2),  161--171 (2021)

\bibitem{2017Quicksilver}
Yang, X., Kwitt, R., Styner, M., Niethammer, M.: Quicksilver: Fast predictive
  image registration--a deep learning approach. NeuroImage  \textbf{158},
  378--396 (2017)

\bibitem{yoo2017ssemnet}
Yoo, I., Hildebrand, D.G., Tobin, W.F., Lee, W.C.A., Jeong, W.K.: ssemnet:
  Serial-section electron microscopy image registration using a spatial
  transformer network with learned features. In: Deep Learning in Medical Image
  Analysis and Multimodal Learning for Clinical Decision Support, pp. 249--257.
  Springer (2017)

\bibitem{2020Recursive}
Zhao, S., Dong, Y., Chang, E.I., Xu, Y., et~al.: Recursive cascaded networks
  for unsupervised medical image registration. In: Proceedings of the IEEE/CVF
  International Conference on Computer Vision. pp. 10600--10610. IEEE (2019)

\end{thebibliography}

\end{document}